\documentclass[sigconf]{acmart}
\usepackage[utf8]{inputenc}

\usepackage{todo}

\usepackage{graphicx}
\usepackage{subfiles}

\usepackage{tabularx}
\usepackage{booktabs}
\PassOptionsToPackage{hyphens}{url}\usepackage{hyperref}

\usepackage{caption}
\usepackage{subcaption}

\AtBeginDocument{%
  \providecommand\BibTeX{{%
    \normalfont B\kern-0.5em{\scshape i\kern-0.25em b}\kern-0.8em\TeX}}}

\copyrightyear{2021}
\acmYear{2021}

\setcopyright{acmlicensed}\acmConference[CIKM '21]{Proceedings of the 30th ACM International Conference on Information and Knowledge Management}{November 1--5, 2021}{Virtual Event, QLD, Australia}
\acmBooktitle{Proceedings of the 30th ACM International Conference on Information and Knowledge Management (CIKM '21), November 1--5, 2021, Virtual Event, QLD, Australia}
\acmPrice{15.00}
\acmDOI{10.1145/3459637.3482415}
\acmISBN{978-1-4503-8446-9/21/11}

\begin{document}

\title{Certifying One-Phase Technology-Assisted Reviews}

\author{David D. Lewis}
\affiliation{%
  \institution{Reveal-Brainspace}
   \city{Chicago}
   \state{IL}
  \country{USA}
}
\email{cikm2021paper@davelewis.com}

\author{Eugene Yang}
\affiliation{%
  \institution{IR Lab, Georgetown University}
  \city{Washington}
  \state{DC}
  \country{USA}
}
\email{eugene@ir.cs.georgetown.edu}

\author{Ophir Frieder}
\affiliation{%
  \institution{IR Lab, Georgetown University}
  \city{Washington}
  \state{DC}
  \country{USA}
}
\email{ophir@ir.cs.georgetown.edu}

\renewcommand{\shortauthors}{Lewis, Yang, and Frieder}

\begin{abstract}
Technology-assisted review (TAR) workflows based on iterative active learning are widely used in document review applications. Most stopping rules for one-phase TAR workflows lack valid statistical guarantees, which has discouraged their use in some legal contexts. Drawing on the theory of quantile estimation, we provide the first broadly applicable and statistically valid sample-based stopping rules for one-phase TAR.  We further show theoretically and empirically that overshooting a recall target, which has been treated as innocuous or desirable in past evaluations of stopping rules, is a major source of excess cost in one-phase TAR workflows. Counterintuitively, incurring a larger sampling cost to reduce excess recall leads to lower total cost in almost all scenarios.                     
\end{abstract}




\maketitle

\section{Introduction}

\textit{Technology-assisted review (TAR)} is the use of technological means to accelerate manual document review workflows. A prominent application is document review in legal cases, known as \textit{electronic discovery} or \textit{eDiscovery} \cite{baron2016perspectives}, a multi-billion dollar industry.\footnote{Global \$12.9 Billion eDiscovery Market Forecast to 2025, \url{https://prn.to/3upSeBC}}  
Another  application area is \textit{systematic reviews} of scientific literature \cite{wallace2010semi}, which have played a revolutionary role in  empirical medicine~\cite{higgins2019cochrane} and other fields~\cite{gough2017introduction}. More generally, TAR is applicable to a range of high recall retrieval tasks \cite{de2013active,li2014req,cuzzocrea2015high,song2017relevance,abualsaud2018system,dias2019trivir}. The TREC-COVID project was an emergency deployment of a TAR process early in the Covid-19 pandemic~\cite{roberts2020trec}.

Two categories of TAR workflows can be distinguished. \textit{Two-phase TAR workflows} (sometimes called \textit{culling} workflows) are focused on iterative training of a text classifier by active learning \cite{settles2009active} which is then used to select a subset of a collection for review \cite{cormack2016scalability, cost-structure-paper, mcdonald2018active}.  A distinction is drawn between the training phase (Phase 1) and the review phase (Phase 2).  While review of documents is done in both phases, most review effort occurs in Phase 2, after training is over. Two-phase reviews are preferred when per-document costs vary among review personnel~\cite{cost-structure-paper}.

In contrast, one-phase workflows do not distinguish between training and review, and are preferable when review costs are constant. Iterative training of models, using those models to prioritize documents for review, reviewing of the prioritized documents, and the feeding back of reviewed documents for training continues during the entire review. This is the structure of a classical relevance feedback workflow in information retrieval \cite{rocchio1971relevance,ruthven2003survey} and, indeed, relevance feedback is widely used in one-phase TAR reviews \cite{cormack2016scalability}.

Since TAR is used when it would be too expensive to review all documents \cite{randreport}, a \textit{stopping rule} is necessary to decide when the review ends.  However, one wants confidence, and ideally a certification by statistical guarantee, that a certain proportion of relevant documents have been found by the stopping point, i.e., that a recall target has been achieved \cite{lewis2016defining}. A stopping rule can thus fail in one of two ways: failing to hit its recall target or incurring unacceptably high costs in doing so.

Unfortunately, no statistically valid, generally applicable stopping rule for one-phase TAR has been available (Section~\ref{sec:other-one-phase-rules}).  The lack of such certification rules has limited the adoption of one-phase TAR workflows. For instance, the US Department of Justice Antitrust Division's model agreement for use of supervised learning includes only two-phase culling workflows.\footnote{\url{https://www.justice.gov/file/1096096/download}}

In response to this need, we reconsider TAR stopping rules from the perspective of statistical quality control in manufacturing~\cite{grant1996statistical}.  Our contributions are: 

\begin{itemize}

    \item A taxonomy of TAR stopping rules by application contexts in which they can be used 

    \item A theoretical framework for understanding stopping a TAR review as a problem in quantile estimation 
    
    \item The first two general purpose certification rules for one-phase TAR: the Quantile Point Estimate Threshold (QPET) rule and the Quantile Binomial Confidence Bound (QBCB) rule. Both can be used with any sample size and recall target. The latter also provides a confidence interval on recall at any specified confidence level.  
    
    \item A theoretical and empirical demonstration that, for many TAR tasks, the counterintuitive key to reducing total TAR review cost is to incur \textit{more} cost for sampling and \textit{reduce} excess recall 
    
\end{itemize}

We begin by proposing a taxonomy of TAR stopping rules and zeroing in on those with broad applicability (Section~\ref{sec:background}). We identify sequential bias as the key challenge to certification rules and apply the theory of quantile estimation to evade this bias (Section~\ref{sec:point-est}). This leads to first the QPET rule (Section~\ref{sec:point-est}) and then the QBCB rule (Section~\ref{sec:confidence-intervals}), whose properties we analyze.  We also examine previously proposed certification rules and find that only one narrowly applicable rule, Cormack and Grossman's Target rule \cite{cormack2014evaluation}, is statistically valid and indeed is a special case of the QBCB rule (Section~\ref{sec:other-sample-rules:target}).  Finally, we demonstrate theoretically and empirically that minimizing sample size, as suggested by Cormack and Grossman, is almost always suboptimal from a total cost standpoint (Sections~\ref{sec:cost-overruns} and \ref{sec:results}).

\section{A Taxonomy of TAR Stopping Rules}
\label{sec:background}

Many TAR stopping rules that have been proposed would be unusable in most operational TAR contexts. In this section, we propose a taxonomy of stopping rules that clarifies their range of applicability. 

TAR evaluation conferences ~\cite{totalrecall2015, totalrecall2016, clef2017ehealth-tar, clef2018ehealth-tar, clef2019ehealth-tar} have emphasized {\em interventional} stopping rules, i.e., rules that alter the method used to select documents for review. These rules include SCAL \cite{cormack2016scalability}, Autostop - Conservative \cite{li2020stop}, Autostop - Optimistic \cite{li2020stop}, and a recent rule by Callaghan and M\"{u}ller-Hansen \citet{callaghan2020statistical}.  By modifying the document selection process, these methods gather information that enables more accurate stopping (if not always valid statistical guarantees).    

While powerful, an interventional rule requires that all documents selected be chosen by a particular novel active learning algorithm. Most document reviews rely on commercial TAR software whose document selection algorithms cannot be modified by the user. Further, review managers often prefer (and may be legally required) to select documents not just by active learning, but also by Boolean text or metadata searches. Documents from other sources (related projects, direct attorney knowledge, or legal actions) may also need to be reviewed at arbitrary times.

In contrast, we call a stopping rule a \textit{standoff} rule if it can be applied to any TAR review, regardless of how documents are selected or in what order. Some rules allow arbitrary review combined with interventional portions: we call these \textit{hybrid} rules.   

Standoff and hybrid rules usually require drawing a random sample for estimation purposes. Some of these rules assume that all review team decisions are correct (\textit{self-evaluation} rules), while others assume only the decisions on the sample are correct (\textit{gold standard} rules).  

A cross-cutting distinction for all rules is how strong a guarantee of quality they provide. \textbf{Heuristic} rules make a stopping decision based on general patterns observed for review processes, such as declining precision with increasing manual search effort or diminishing impact from new training data \cite{saha2015batch-mode, cormack2014evaluation, quantstop-paper, cormack2016engineering, wallace2010semi}.  Heuristic stopping rules for one-phase TAR reviews are closely related to stopping rules for active learning in two-phase TAR reviews \cite{mcdonald2018active} and in generalization tasks \cite{settles2009active, kottke2019limitations, tomanek2008approximating}.

\textit{Certification} rules, on the other hand, use a random sample to provide a formal statistical guarantee that the stopping point has certain properties and/or provide a formal statistical estimate of effectiveness at the stopping point. If correctly designed, they give a degree of confidence that heuristic rules cannot. However, with one narrow exception, previously proposed certification rules fail to meet their purported statistical guarantees (Section~\ref{sec:other-one-phase-rules}). 

The consequences for such failures can be severe: parties in legal cases have been sanctioned for failing to meet stated targets on information retrieval effectiveness measures.\footnote{\textit{In Re: Domestic Airline Travel Antitrust Litigation, 1:15-mc-01404 (D.D.C. Sept. 13, 2018)}}. In Sections~\ref{sec:point-est} and \ref{sec:confidence-intervals} we provide the first standoff gold standard certification rules for one-phase TAR workflows that can be used with any sample size, recall target, and confidence level.

\section{A Point Estimation Rule}
\label{sec:point-est}

Certification rules condition stopping on some statistical guarantee of effectiveness of the TAR process. We consider here the usual collection-level binary contingency table measures, where the four outcomes TP (true positives), FP (false positives), FN (false negatives), and TN (true negatives) sum to the number of documents in the collection. For a one-phase TAR workflow, a positive prediction or detection corresponds to the document having been reviewed before the workflow is stopped.

Recall, $TP/(TP+FN)$, is the most common measure on which the TAR processes are evaluated \cite{lewis2016defining}.
Other measures of interest in TAR are precision $= TP/(TP+FP)$ and elusion $= FN / (FN + TN)$. Elusion (which one desires to be low) can be thought of as precision in the unreviewed documents and has mostly seen use in the law \cite{roitblat2007search}.

\subsection{Estimates and Estimators}

Effectiveness must be estimated.  Estimates are produced by \textit{estimators}, i.e., functions that define a random variable in terms of a random sample from a population \cite{lehmann2006theory}. An estimate is the value taken on by that random variable for a particular random sample. A \textit{point estimate} is an estimate which is a scalar value. A common point estimator is the \textit{plug-in estimator}, which replaces population values by the random variables for the corresponding sample values \cite{flury2013first}. The plug-in estimator for recall, based on a simple random sample annotated for both category and detection status, is $X/Y$ where $X$ is a random variable for the number of positive detected examples in the sample, and $Y$ is a random variable for the total number of positive examples in that sample. In other words, recall on the labeled random sample is used as the point estimate of recall in the population.

The plug-in estimator of recall assumes that both class labels and detection statuses are known. If we are using the estimate in a stopping rule, however, we must stop to have an estimate, but must have an estimate to stop. The usual resolution of this dilemma in TAR is to compute, after each batch of documents is reviewed, what the estimated effectiveness would be if the TAR process were stopped at that point. The TAR process is stopped the first time one of these trial estimates exceeds the recall goal. We refer to this rule, widely used in practice, as the \textit{Point Estimate Threshold (PET)} stopping rule.

\subsection{The PET Rule is Invalid}

Unfortunately, the PET rule is statistically biased: the expected value of effectiveness at the stopping point typically falls short of the claimed effectiveness level. We demonstrate this with an example that, while simple,  exhibits the core phenomena at play.  

Consider a large collection, A, with an even number $N$ of documents, \textit{all} of which are relevant. If we ran the TAR process until all documents were found, each document would be assigned a rank corresponding to the order in which it was found. Call that number the A-rank of the document. Suppose our recall goal is 0.5. Since all documents are relevant in our example, a TAR process achieves recall of 0.5 or more if it stops at or after A-rank $N/2$.  

Now draw a simple random sample of even size $n$ from A and have it coded, prior to starting the TAR process, as a gold standard. On coding those documents we will find all are relevant, and so we have a simple random sample, $D$, of size $r = n$ from the relevant documents in A. At first we do not know the A-rank of any document in sample D. However, as reviewers examine documents, they periodically find one of the $n$ sample documents, at which point we know its A-rank. When the $n/2$'th document from the sample is found, the plug-in estimate of recall on the sample will be 0.5, and the PET rule would stop the TAR process.

Since D is a random sample, the value of the $n/2$'th highest A-rank in D, our stopping point, is a random variable $D_{(n/2)}$, the $n/2$'th \textit{order statistic} in sample $D$ \cite{ABN}. It has the following probability mass function 

\begin{equation}
    f_{(n/2):n}(a) 
         = \frac{
                  \binom{a-1}{(n/2) - 1} \binom{1}{1} \binom{N-a}{(n/2)}
                }
                {
                  \binom{N}{n}
                },
\end{equation}

\noindent
corresponding to $n$ draws without replacement from three bins (less than, equal to, and greater than $a$) \cite[Chapter~3]{ABN}. The expected value of $D_{(n/2)}$ is $(n/2)(N+1)/(n+1)$ \cite[Chapter~3]{ABN}, and thus the expected recall of the PET rule in this case is $(1/2)(n/(n+1))((N+1)/N)$. This is less than 0.5 for any $r < N$.

\subsection{Quantile Point Estimation}

The PET rule makes multiple tests on estimates and stops when the first one succeeds, thus biasing the estimate at the stopping point. This phenomenon is the focus of sequential analysis \cite{siegmund2013sequential, wald2004sequential, dimm2017confirming, webber2013sequential}, which is central to statistical quality control in manufacturing \cite{grant1996statistical}. A key insight from sequential analysis is that conditioning stopping of a process on a random variable makes the stopping point itself a random variable. It is that latter random variable we need to have the desired statistical properties. 

Suppose we view the PET rule more abstractly, as a rule that stops a TAR process when we have found $j$ items from a sample D of positive documents from A. The A-rank of the $j$th item will be the $j$th lowest A-rank in our sample. That value, $d_{j}$, is the realization for our sample of the random variable $D_{j}$, where $D_{1},...,D_{r}$ are the order statistics for the sample \cite{ABN}.  

For such a rule to be valid, given a recall goal $t$ and positive sample size $r$, one strategy would be to choose $j$ such that the worst case expected value of recall for any data set, averaged over the realizations $d_{j}$ of $D_{j}$ for that data set, is at least $t$. 
Computing this worst case expected value is nontrivial. 
Fortunately an alternative perspective is possible when recall is the measure of interest.

A \textit{t-quantile} is a population parameter such that a fraction $t$ of the population is at or above that value. Formally, for $0 < t < 1$, we define $b_{t}$ to be the $t$-quantile of finite population B if $b_{t} = \inf\{x: P[X \le x] \ge t\}$ for X drawn uniformly from the population \cite[Chapter~7]{david2003order}. Let B be just the relevant documents within collection A, but in sorted order by their A-ranks. Let A-rank $b_{t}$ be the $t$-quantile for B. Then the recall of a TAR process stopping at $b_{t}$ is the smallest $t'$ such that $t' \ge t$ and $t'$ is achievable for some stopping point.

\begin{figure}
    \centering
    \includegraphics[width=\linewidth]{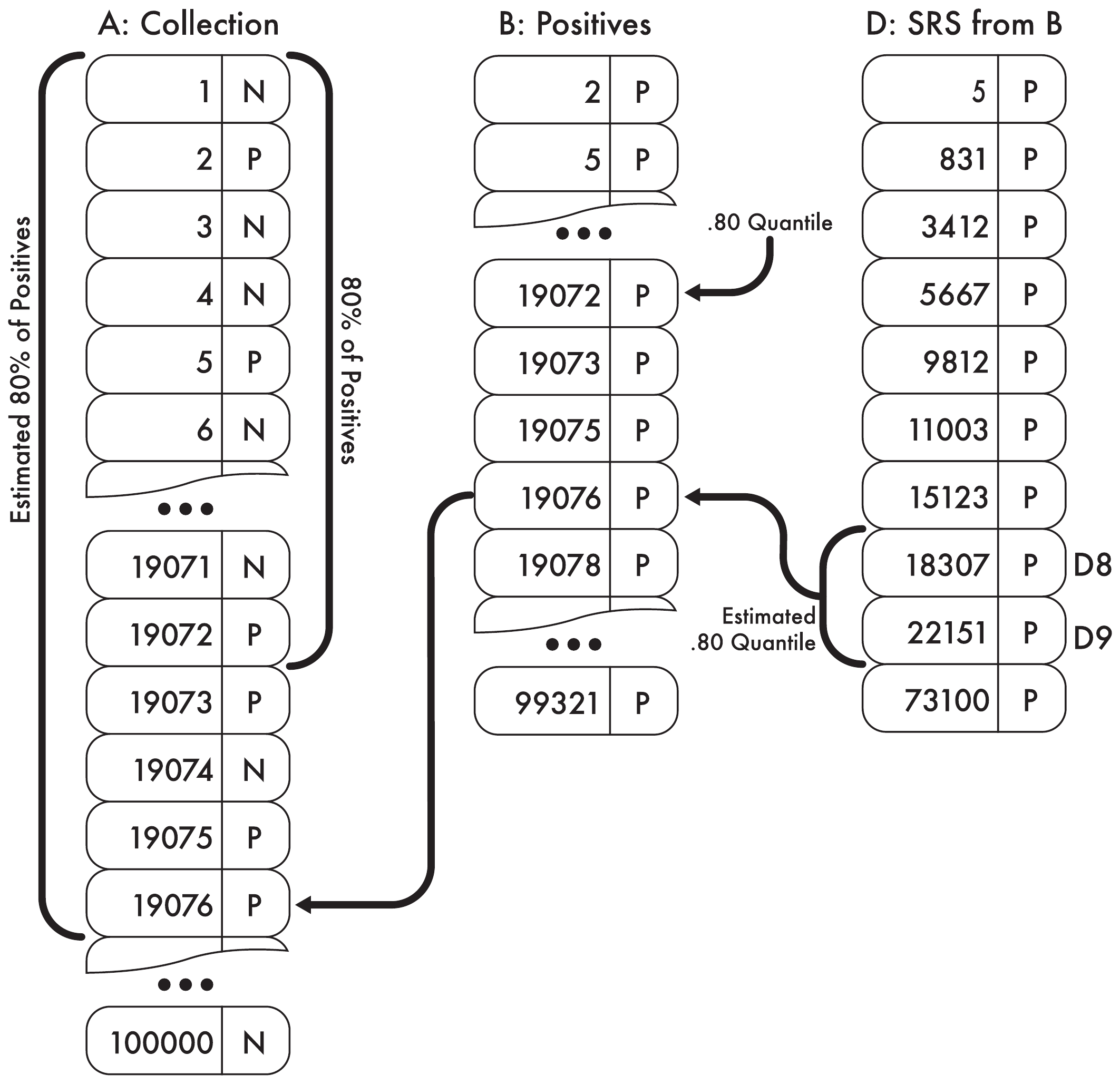}
    \caption{\textit{A quantile point estimation example.} A TAR process if carried to the end assigns an A-rank to each document in a collection (A), and thus to each document in the positive subpopulation (B). In this example, A-rank 19072 is the 0.80 quantile in B, meaning that a TAR process stopping at that rank has 0.80 recall.  D is a simple random sample of size 10 from B. Hyndman \& Fan's Q7 point estimator of the 0.80-quantile is thus based on the order statistics $D_{8}$ and $D_{9}$. Here the realizations are 18307 and 22151, leading to 19076 (slightly higher than the true value) being the point estimate of the 0.8-quantile. (Diagram by Tony Dunnigan.)}
    \label{fig:quatile-esitmation-diagram}
\end{figure}

The quantile perspective links recall at a stopping point to a single population parameter. We then require an estimator that maps from order statistics $D_{j}$ in the positive sample D to population quantiles within the positive subpopulation B, i.e. a \textit{quantile point estimator}. Hyndman \& Fan \cite{hyndman1996sample} review the properties of nine quantile point estimators, of which their Q7 is the default in the R statistical package.\footnote{https://www.rdocumentation.org/packages/stats/versions/3.6.2/topics/quantile} Q7 is defined by letting $h = (r - 1)t + 1$, $j = \lfloor h \rfloor$, and using $D_{j} + (h-j)(D_{j+1} - D_{j})$ as the estimator for the $t$-quantile. Figure~\ref{fig:quatile-esitmation-diagram} diagrams the logic of quantile estimation using Q7.

Using Q7 as our quantile point estimator, we define the \textit{Quantile Point Estimate Threshold (QPET)} stopping rule as follows. Given a sample size $r$ and recall goal $t$, we compute $h = (r - 1)t + 1$ and $j = \lfloor h \rfloor$. We need to stop at a point where we can apply our estimator $D_{j} + (h-j)(D_{j+1} - D_{j})$. If $h$ is an integer, then $j=h$ and we only need the value of $D_{j}$. We therefore stop at $d_{j}$, i.e., after finding the $j$'th positive sample document. If $h$ is not an integer, then we need the values of both $D_{j}$ and $D_{j+1}$, so we stop at $d_{j+1}$, i.e., after finding the $j+1$'th positive sample document.   In either scenario, $d_{j} + (h-j)(d_{j+1} - d_{j})$ is our point estimate of the $t-$quantile at the stopping point (with the second term 0 if $h$ is an integer). 

If a point estimate of recall at the stopping point is required, we can use $t$ as that estimate. This point estimate of recall is conservative in the sense that recall at the $t$-quantile is always at least $t$, but can be higher. 

When does the QPET rule stop in comparison with the PET rule?  Assume a nontrivial recall goal $0 < t < 1$ and positive sample size $r$. The PET rule stops after reviewing $d_{j}$ total examples, where $j$ is lowest value such that $j/r \geq t$. If $rt$ is an integer, this will be when $j = rt$. In this case, the QPET rule has $j' = \lfloor h \rfloor =  \lfloor (r - 1)t + 1 \rfloor = \lfloor rt + 1 - t \rfloor = rt + \lfloor 1 - t \rfloor = rt = j$.  Since $t = j/r$, we can also write $h = (r-1)t + 1 = j(r-1)/r + 1$. Ignoring the trivial case $r=2$, $(r-1)/r$ is never an integer, and thus $h$ is never an integer. So both $D_{j'}$ and $D_{j'+1}$ are needed, and QPET stops at $d_{j+1}$.  

If $rt$ is not an integer, then the PET rule stops at $d_{j}$ where $j = \lceil rt \rceil = \lfloor rt \rfloor + 1$. Suppose first that $(r-1)t$ \textit{is} an integer. Then for the QPET rule $j' = \lfloor (r-1)t + 1 \rfloor = (r-1)t + 1 = rt + (1-t)$. Since $0 < (1-t) < 1$, this is either $\lceil rt \rceil$ or $\lceil rt \rceil + 1$.  Conversely, if $(r-1)t$ is not an integer, $j' = \lfloor (r-1)t + 1 \rfloor = \lfloor rt + (1-t) \rfloor$, which is either $\lfloor rt \rfloor$ or $\lfloor rt \rfloor + 1$.  

In all cases then, the QPET rule requires finding at most one more positive sample document than the PET rule, and sometimes, no additional documents.  
  
\section{A Confidence Interval Rule}
\label{sec:confidence-intervals}

The QPET rule outputs a point estimate on the $t$-quantile (i.e., the number of documents to review), and with a point estimate of recall equal to $t$. However, we surely should feel less good about these estimates if they are based, say, on a sample of 10 positive examples than on a sample of 1000 positive examples. 

A {\em confidence interval} is an estimate that consists of a pair of scalar values with an associated confidence level, conventionally expressed as $1 - \alpha$ \cite[Chapter~2]{hahn2011statistical}. A confidence interval estimator specifies a closed interval $[X,Y]$, where $X$ and $Y$ are random variables defined in terms of the sample. We say that such an estimator produces a  $1 - \alpha$ confidence interval for a population value $z$ when the probability is at least $1 - \alpha$ that, over many draws of a random sample of the specified type, the sample-based realizations $x$ of $X$ and $y$ of $Y$ are such that $x \le z \le y$.

Confidence interval estimators are used in two ways in TAR. The first is to use a power analysis on the estimator as a guide to sample size \cite{ryan2013sample}.  A review manager will 
draw a random sample large enough to guarantee that any confidence interval estimate produced from it will have some property, such as a maximum margin of error of 0.05.

The second use of confidence intervals is in reporting, i.e., in making a statistical claim based on a labeled random sample about the effectiveness achieved by a TAR process. If the only use of the random sample is in reporting, this is unproblematic. But if (as is common) the same sample is used to decide when to stop the review, the reported confidence interval estimate will have sequential bias \cite{webber2013sequential}.

\subsection{Quantile Confidence Intervals}

As with point estimates, the quantile perspective can rescue confidence intervals on recall from sequential bias.  A \textit{quantile confidence interval} is a confidence interval on a quantile \cite[Chapter~4]{wilcox2011introduction}.  To avoid distributional assumptions about the TAR process, we can use a nonparametric quantile confidence interval estimator \cite[Chapter~7]{david2003order}.  This takes the form of a pair of order statistics, $[D_{i}, D_{j}]$. The estimator determines the values $i$ and $j$ based on the quantile level $t$, sample size $r$, and confidence level $1 - \alpha$. It provides the guarantee that, with at least $1 - \alpha$ probability over draws of the random sample, the $t$-quantile $b_t$ falls within the sample-based realization $[d_{i}, d_{j}]$.

If an estimator of this form is available, we can define the stopping point of a TAR review to be the value of $d_{j}$ for our positive random sample, and have $1-\alpha$ confidence that the $t$-quantile in B falls within $[d_{i}, d_{j}]$. By definition of the $t$-quantile, we thus have $1 - \alpha$ confidence that stopping at $d_{j}$ gives a recall of at least $t$. 

For many uses of confidence intervals we want estimators that make the interval narrow (the realization $d_{j} - d_{i}$ is likely to be small) and/or symmetric (i.e., $d_{j} - b_t$ and $b_t - d_{i}$ are likely to be similar, where $b_{t}$ is some point estimate). For a stopping rule, however, the most important criterion is that $d_{j}$ is likely to be small, since this reduces the number of documents the TAR process must review before stopping. 

We can minimize the likely value of $d_{j}$ by using a nonparametric \textit{one-sided upper confidence interval (UCI)} on a quantile \cite[Chapter~5]{hahn2011statistical}. Such an interval has the form $[D_{0}, D_{j}]$, where $D_{0}$ is the $0$th order statistic, i.e., the lowest logically possible value. For us this is $D_{0} = 1$ (the lowest A-rank); so the interval is $[1, D_{j}]$. We refer to the pair as an \textit{1-s UCI}, and the upper end of the interval $D_{j}$ as a \textit{1-s UCB} (\textit{one-sided upper confidence bound}).    

The estimator must choose $j$ such that the realization $d_{j}$ will be, with $1-\alpha$ probability, a $t$-quantile or higher. This is equivalent to requiring a probability $1 - \alpha$ or higher that fewer than $j$ elements of positive random sample D have A-rank less than the $t-$quantile. Suppose there are $R$ positives in B, and that our sample of positives is of size $r$. Then our estimator should choose the smallest $j$ such that:  

\begin{equation}
    \sum_{k=0}^{j-1}
       \frac{
             \binom{\lceil tR \rceil - 1}{k}  
             \binom{R - \lceil tR \rceil + 1}{r - k}
            }
            {
               \binom{R}{r}
            }
         \ge 1 - \alpha
\end{equation}

In a TAR setting we do not know $R$. However, if $R$ is large relative to $r$, the binomial distribution is a good approximation to the above hypergeometric distribution \cite[Chapter~3]{thompson1997theory}. In this condition, we want the smallest $j$ such that 
\begin{equation}
    \sum_{k=0}^{j-1} \binom{r}{k} t^{k} (1-t)^{r-k} \ge 1 - \alpha. 
    \label{eq:binomial-estimation}
\end{equation}

In fact, we can use the binomial approximation safely even when we are not confident that $r$ is small relative to $R$. For values of $t$ greater than 0.5, the fact that the binomial has larger variance than the hypergeometric means that the $j$ chosen using the binomial approximation will never be less than the one chosen using the hypergeometric. Values of recall less than 0.5 are rarely of interest, but if needed we could find a similarly conservative value of $j$ for such a $t$ by running the summation from $r$ downwards instead of $0$ upwards.  

Based on the above analysis, we define the \textit{Quantile Binomial Confidence Bound (QBCB)} stopping rule.  Given a sample size $r$, recall target $t$, and confidence level $1 - \alpha$, it specifies stopping a one-phase TAR process when the $j$th positive sample document is found. Here $j$ is smallest integer such that $[1, D_{j}]$ contains the $t$-quantile from the unknown population of positive examples with $(1-\alpha)$ confidence, based on the binomial approximation.  

\subsection{The QBCB Rule and Recall Estimation}

We observed that the recall goal $t$ can be used as a conservative point estimate of recall at the QPET stopping point. By the same logic, $t$ is a conservative point estimate of recall at the QBCB stopping point. 

If we prefer an interval estimate, we can use a $1 - \alpha$ \textit{one-sided lower confidence interval (1-s LCI)} (or \textit{one-sided lower confidence bound, 1-s LCB}) estimator \cite[Chapter~2]{hahn2011statistical}). This defines a pair $[L, 1.0]$ where, with probability at least $1-\alpha$ over random samples of size $r$, the realization $[\ell, 1.0]$ contains a desired population value. Given the definition of $t$-quantile, we know that $[t, 1.0]$ is a $1-alpha$ 1-s LCI on recall at the QBCB stopping point. 

This interval estimate on recall may seem unsatisfying: it is identical regardless of sample size. However, this simply reflects the task we have set for the QBCB rule: stop as soon as one has confidence that a recall goal has been met. Larger sample sizes translate to earlier stopping, not a tighter 1-s LCI.

\subsection{What to Expect from the QBCB Rule}

We can also compute more conventional estimates of recall at the QBCB stopping point.  As long as those estimates depend only on $j$ (which is fixed as soon as we choose $r$ and $t$) and not on $d_j$ (the actual A-rank at which we stop), these estimates are not affected by sequential bias. These estimates give insight into the behavior of the QBCB rule.

Table~\ref{tab:quantile-estimation-sample-sizes} shows the QBCB values of $j$ for recall goal 0.8, confidence level 95\% ($1 - \alpha = 0.95$), and selected sample sizes $r$ from 14 to 457. (The choice of the sample sizes is discussed in Section~\ref{sec:cost-overruns}.)  

Sample sizes 8 to 13 are also included. However, with these sample sizes, the only 95\% 1-s UCI based on order statistics that includes the 0.8-quantile is the trivial interval $[D_{0},D_{r+1}]$ = $[1,N]$ (using the convention that the $r+1$'th order statistic for a sample of size $r$ is the maximum population value). So for $r < 14$, the QBCB value of $j$ is $r+1$, and the rule does not provide meaningful stopping behavior.  For these sizes we instead show $j^{*} = j-1 = r$, the largest non-trivial stopping point. We also show both the QBCB $j$ and $j^{*} = j-1$ for the case r=21, discussed in Section~\ref{sec:cost-overruns}. Rows with QBCB $j^{*} = j-1$ values are indicated by "*".

We show the value of three estimates of recall based solely on $j$ or $j^{*}$. The first is a 95\% 1-s LCI, but for recall rather than for the  $t$-quantile. In particular, we use the Clopper-Pearson exact interval \cite{brown2001interval}. Second, we show the plug-in estimate $j/r$ discussed earlier for the PET rule.  Finally, we show a 95\% 1-s UCI on recall, again computed using the Clopper-Pearson method. 

For rows with the QBCB $j$ value, the lower end of the 95\% 1-s LCI is always at or above 0.80, but fluctuates and is closer to 0.80 when the sample size is larger.  This reflects the fact that the Clopper-Pearson LCI is based on the same binomial approximation used in the QBCB rule. The only difference is that the QBCB computation solves for integer $j$ based on fixed real $t$, while the LCI computation solves for real $t$ based on fixed integer $j$. The QBCB requirement that $j$ be an integer means that the $t$ at the lower end of the LCI is typically slightly more than 0.8, with the difference decreasing as $r$ increases and more $j$ values are available. 

The plug-in point estimates (which are simply $j/r$ or $j^{*}/r$ depending on the row) for small sample sizes are much higher than 0.8. 
We can think of these as the estimated recall at which the naive PET rule would need to stop to achieve the same confidence bound as the QBCB rule, and reflects how uncertain recall estimates from small samples are.

The last column shows a 95\% 1-s UCI on recall at the QBCB stopping point. This estimate shows that, as sample sizes increase, we slowly become more confident that the QBCB stopping point will not have very high recall. Section~\ref{sec:cost-overruns} discusses why, counterintuitively, we should want such confidence.

\begin{table}
\centering
\caption{Stopping points within a positive sample for the QBCB rule with 0.80 recall and 95\% confidence estimates, plus three conventional estimates of recall at the stopping point.}
\label{tab:quantile-estimation-sample-sizes}
\begin{tabular}{|c|c|c|c|c|c|}
 \hline
           &              & \multicolumn{3}{c|}{Recall Estimators Applied at j} \\
 $r$       &  $j$         & 95\% 1-s LCI   & Plug-in  & 95\% 1-s UCI  \\
  \hline 
  8        &   8*          & [0.688, 1.000] & 1.000    & [0.000, 1.000] \\
  9        &   9*          & [0.717, 1.000] & 1.000    & [0.000, 1.000] \\
 10        &  10*          & [0.741, 1.000] & 1.000    & [0.000, 1.000] \\
 11        &  11*          & [0.762, 1.000] & 1.000    & [0.000, 1.000] \\ 
 12        &  12*          & [0.779, 1.000] & 1.000    & [0.000, 1.000] \\
 13        &  13*          & [0.794, 1.000] & 1.000    & [0.000, 1.000] \\
  \hline                         
 14        &  14          & [0.807, 1.000] & 1.000    & [0.000, 1.000] \\
 21        &  20*         & [0.793, 1.000] & 0.952    & [0.000, 0.998] \\
 21        &  21          & [0.867, 1.000] & 1.000    & [0.000, 1.000] \\
 22        &  21          & [0.802, 1.000] & 0.955    & [0.000, 0.998] \\
 29        &  28          & [0.847, 1.000] & 0.965    & [0.000, 0.998] \\ 
 30        &  28          & [0.805, 1.000] & 0.933    & [0.000, 0.988] \\ 
 31        &  29          & [0.811, 1.000] & 0.936    & [0.000, 0.988] \\
 37        &  34          & [0.804, 1.000] & 0.912    & [0.000, 0.978] \\
 44        &  40          & [0.804, 1.000] & 0.909    & [0.000, 0.968] \\
 50        &  45          & [0.801, 1.000] & 0.900    & [0.000, 0.960] \\
 63        &  56          & [0.801, 1.000] & 0.889    & [0.000, 0.947] \\
 76        &  67          & [0.803, 1.000] & 0.882    & [0.000, 0.937] \\
 88        &  77          & [0.802, 1.000] & 0.875    & [0.000, 0.928] \\
 106       &  92          & [0.801, 1.000] & 0.868    & [0.000, 0.918] \\
 129       &  111         & [0.800, 1.000] & 0.861    & [0.000, 0.908] \\
 158       &  135         & [0.800, 1.000] & 0.854    & [0.000, 0.898] \\
 198       &  168         & [0.800, 1.000] & 0.849    & [0.000, 0.889] \\
 255       &  215         & [0.800, 1.000] & 0.843    & [0.000, 0.879] \\
 332       &  278         & [0.800, 1.000] & 0.837    & [0.000, 0.870]\\
 457       &  380         & [0.800, 1.000] & 0.832    & [0.000, 0.860]\\
\hline
\end{tabular}    

\end{table}

\section{Sample Size and Recall}
\label{sec:cost-overruns}

Past evaluations of stopping rules have often treated overshooting a recall goal
as a lucky outcome~\cite{cormack2016engineering}.  By definition, however, a certification rule that stops with an recall higher than its goal has incurred extra costs.  A TAR process that incurs high costs, particularly unpredictably high costs, while overshooting stakeholder requirements is not a success. 

Further, in some contexts exceeding a recall goal may be a negative outcome even if costs are ignored. A litigant that would like to produce 0\% of responsive documents to an adversary, but has a legal obligation to produce 80\% of responsive documents, is not happier if their legal service provider delivers 90\% of responsive documents to the adversary.

Recall is an expensive measure on which to overshoot a goal. As a TAR method pushes for high recall, relevant documents tend to be spaced increasingly farther apart.  This is a basis of the common heuristic rule that stops review when batch precision drops below some minimum value.  Larger intervals between relevant documents mean that each percentage point of recall achieved beyond the goal value comes at increasing marginal cost. 

Thus part of the benefit of using a larger random sample in a certification rule is \textit{lower} recall. Indeed, jointly choosing an order statistic and a sample size so that both a UCB and an LCB are bounded is an old technique from statistical quality control \cite{guenther1972tolerance}.  

For Table~\ref{tab:quantile-estimation-sample-sizes} we chose the sample sizes $r \ge 30$ to be the smallest sizes for which the 95\% 1-s LCB on recall is less than or equal to each of the values 0.99 to 0.86, decreasing by increments of 0.01. For instance, 158 is the smallest sample size such that the 95\% 1-s LCB on recall is 0.90 or lower.  For sample sizes of 14 and above we always have 95\% confidence that we achieve the specified minimum recall, 0.80.  What we get for larger sample sizes is a lower expected recall (point estimate) and, as shown by the 1-s LCI column, confidence that we will not stop with very high (and expensive) recall.  

For small sample sizes, an additional consideration arises. Consider the first sample size for which we are able to leave $r-j$ sample examples undetected and still hit the desired LCB on recall. As the examples (21,20), (21,21), and (22, 21) show, $r=22$ is the lowest sample size for which $r-j = 1$, i.e., we can leave one example undetected and still meet our criterion. For sample sizes from $r=23$ through $r=29$ the LCB, point estimate, and UCB of recall all increase steadily with increasing sample size, with the largest values at $r=29$. This is a lose-lose situation: increasing sample size in this range both increases sampling costs and increases TAR costs (since we expect to stop at a higher recall). The pattern is not broken until $r=30$, the lowest sample size for which we can leave two examples undetected, at which point the pattern starts again. 

This pattern results from the fact that a sample of size $r$ only provides $r+1$ possible stopping points if stopping is at an order statistic.  Some combinations of sample size, confidence level, and population parameter (recall goal) inevitably poorly match the available choices. This problem decreases for larger sample sizes, since more order statistics are available. As in other estimation situations with small sample sizes, careful choice of sample size can reduce costs substantially \cite[Chapter~2]{ryan2013sample}. 

This phenomenon is also relevant to empirical studies of certification rules: poor choices of sample size will introduce unneeded variation in the relationship between sample size and achieved recall (and thus cost). In our tests in Section~\ref{sec:results} we use the optimal sample sizes from Table~\ref{tab:quantile-estimation-sample-sizes}. 

For the most part, however, larger samples reduce excess recall.  How large a sample is appropriate depends on how much overshooting the recall goal costs. This depends on numerous details, including the difficulty of the classification problem, size of the collection, type of classifier, active learning approach, and batch size.  In Section~\ref{sec:results}, we examine some typical situations.

\section{Proposed Certification Rules}
\label{sec:other-one-phase-rules}

We previously discussed the PET rule and our proposed QPET and QBCB rules. In this section, we examine other certification stopping rules in common TAR practice or proposed in the scientific literature.

\subsection{Repeated PET Rules}

Practitioners often carry out a one-phase TAR workflow until a heuristic rule suggests that they have found most relevant documents. A common hybrid stopping approach is to first do this, then draw a random sample from the unreviewed documents, and make some statistical test on this sample. If the test succeeds, review stops.  If the test fails, the sample is recycled as training data, and the review is restarted until the heuristic again indicates stopping and sampling.  This can be thought of as a repeated PET (RPET) rule: we repeatedly test against some threshold value until succeeding. 

One statistical test used is accept on zero \cite{roitblat2013measurement,hahn1974minimum,newman2018alternative}, i.e., recycle unless no relevant documents are in the sample. More generally one can estimate elusion from the sample, and recycle unless elusion is low enough. A variant on this uses the elusion estimate to compute an ad hoc estimate of recall~\cite{tredennick2015tar}, and recycles unless estimated recall is high enough.  Regardless of the particular variant, all RPET approaches suffer from sequential bias induced by multiple testing: the process is more likely to stop when sampling fluctuation gives an over-optimistic estimate of effectiveness. Dimm~\cite{dimm2017confirming} provides a detailed analysis of how accept on zero fails when used in an RPET rule.

\subsection{The Countdown Rule}

Shemilt, et. al. discuss systematic review projects in which several stopping criteria were considered \cite{shemilt2014pinpointing}.  One is based on what they call the BIR (Baseline Inclusion Rate): simply the plug-in estimate $\hat{p} = r/n$ of the proportion of relevant documents in the collection. They convert this to an estimate $(Nr)/n$ of the number of relevant documents in the collection. They propose stopping the TAR process when the number of relevant documents found equals this value, or the budgeted time runs out.  This is equivalent to using $(R_{a}n)/(Nr)$ as an estimator for recall, and stopping when estimated recall hits a recall target $t$, which for Shemilt was $1.0$.

This stopping rule is known in e-discovery as the  ``countdown method'' or ``indirect method''.\footnote{https://www.courtlistener.com/docket/4259682/304/kleen-products-llc-v-international-paper/} The method is seriously flawed. First, the countdown estimator can produce recall estimates greater than 1.0.  Second, in those cases where the point estimate of the number of relevant in the population is an overestimate, the TAR process may reach the end of the collection without stopping. Finally, the countdown method does not take into account sampling variation, and so provides no statistical characterization of the actual recall achieved.

\subsection{The Target Rule}
\label{sec:other-sample-rules:target}

The Target rule \cite{cormack2016engineering} uses a simple random sample of 10 positive examples (the target set) and stops when the one-phase TAR process has found all of them.
It would be viewed in our framework as implicitly computing a 1-s UCI $[1, D_{10}]$ based on a positive sample of size 10, and stopping when the realization of $D_{10}$ is reached.   

Cormack and Grossman analyze the Target rule and conclude it achieves a recall of 0.70 with 95\% confidence. However, their analysis uses the binomial approximation in an unnecessarily conservative way, by treating $0.3R/R = 0.3$ as small. In fact, Table~\ref{tab:quantile-estimation-sample-sizes} shows that a target set of only 9 positive documents is sufficient to achieve a recall goal of 0.70 with 95\% confidence, while their suggested target set of 10 positive documents achieves a recall goal slightly over 0.74.

The Target rule satisfies (actually exceeds) its claimed statistical guarantee, but does not allow any flexibility in recall goal or confidence level. Further, as shown in  Section~\ref{sec:results}, using the minimum possible positive sample size usually increases total review cost. Requiring that every positive sample document be found also means a single coding error would have large consequences.

\section{Experiment: Methods}
\label{sec:methods}

The correctness of the QPET and QBCB stopping rules is completely determined by the theory of quantile statistics, regardless of sample size. Our goal in empirical work here is not, therefore, to verify the correctness of the rules, but simply to provide a demonstration of how sample size and cost interact in perhaps counterintuitive ways.

We worked with a random 20\% subset of the RCV1-v2~\cite{rcv1} text categorization collection defined in a prior TAR study~\cite{cost-structure-paper}.  An advantage of RCV1-v2 over collections used in past TAR evaluations is the ability to explore a range of category difficulties and prevalences simultaneously.  That study defined three levels of category prevalence and three of classification difficulty. For our demonstration, we selected the category with closest to median difficulty and prevalence from each of their nine bins, and seed document with closest to median difficulty for each category.
Based on that seed document, iterative relevance feedback with a batch size of 200 was carried out until the collection was exhausted (805 iterations).  Supervised learning used the logistic regression implementation in \texttt{scikit-learn} with L2 regularization and 1.0 as the penalty strength.

The resulting batches were concatenated in order.  When applying the QBCB rule we considered stopping points only at the end of each batch, so order within bins had no effect.  For each category and each positive sample size value, we then generated 100 simple random samples constrained to have exactly that number of positive examples.  We applied the QBCB rule with 95\% confidence and recall goal 0.80 to those samples, found the stopping iteration, and computed actual recall and cost at that point. Sample sizes used were all those from Table~\ref{tab:quantile-estimation-sample-sizes} that allow the confidence level and recall goal to be met. 

We separated the review cost at a stopping point into four components for analysis purposes: the positive and negative documents in the random sample, and the positive and negative documents encountered during relevance feedback prior to the stopping point. We assume that the random sample is, to avoid bias, reviewed by different personnel than conduct the main review. Thus encountering the same document in both the sample and during relevance feedback costs twice.  We discuss costs further in the next section.

\section{Experiment: Results and Analysis}\label{sec:results}

\begin{figure}
    \centering
    \includegraphics[width=\linewidth]{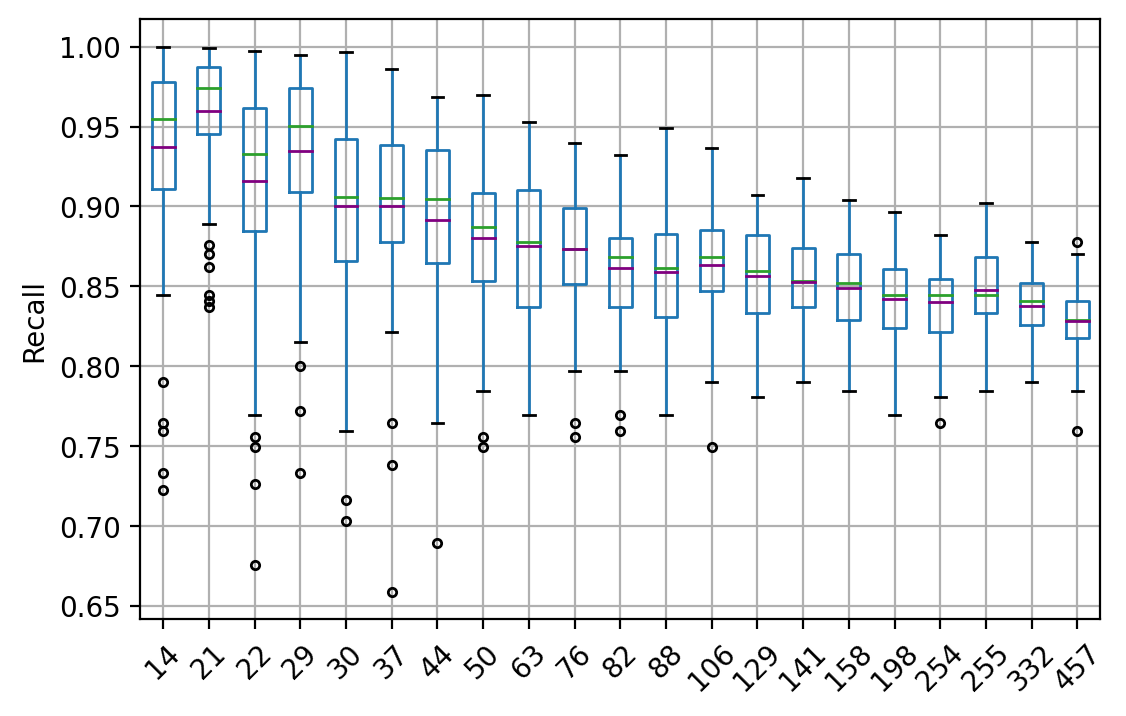}
    \caption{Relationship between positive sample sizes (x-axis) and collection recall at the stopping point (y-axis) for the QBCB rule on category \texttt{E12}. 100 replications of each sample size are displayed using boxplot conventions: the box ranges from the 25\% (q1) to 75\% (q3) quartiles of recall with the 100 replications, the green and purple lines are the median and mean recall respectively, and whiskers extend to the lesser of the most extreme cost observed or to a distance 1.5(q3 - q1) from the edges of the box. Outliers are presented as dots above and below the whiskers.}
    \vspace{-1.5em}
    \label{fig:recall-boxplot}
\end{figure}

\begin{figure*}
    \centering
    \includegraphics[width=\linewidth]{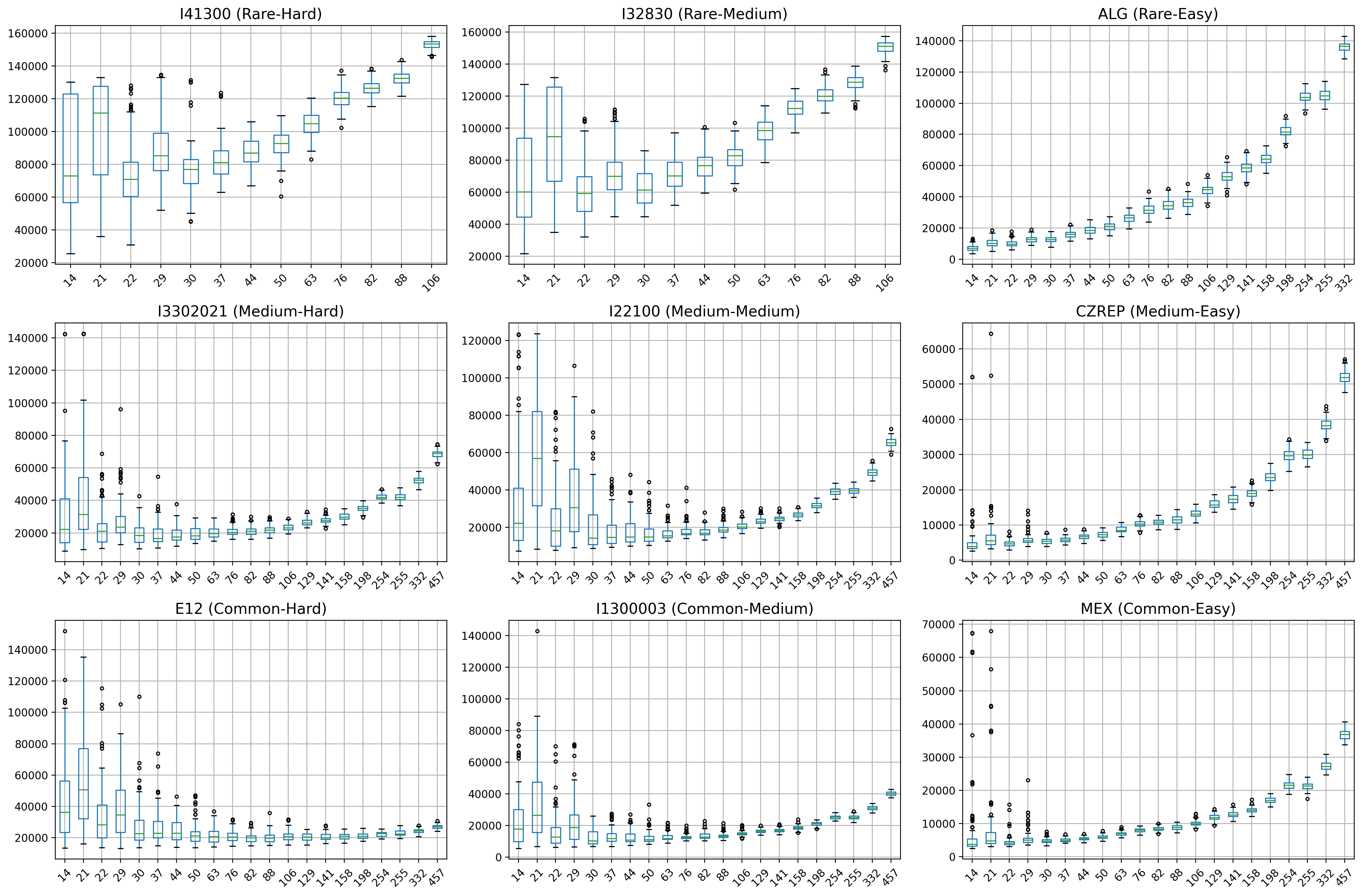}
    \caption{Relationship between positive sample sizes (x-axis) and total review cost (including sampling) at the QBCB stopping point (y-axis). 100 replications of each sample size are used.  Boxplot conventions are as in Figure 2.}
    \label{fig:cost-boxplot}
\end{figure*}

\begin{figure*}
    \centering
    \includegraphics[width=\linewidth]{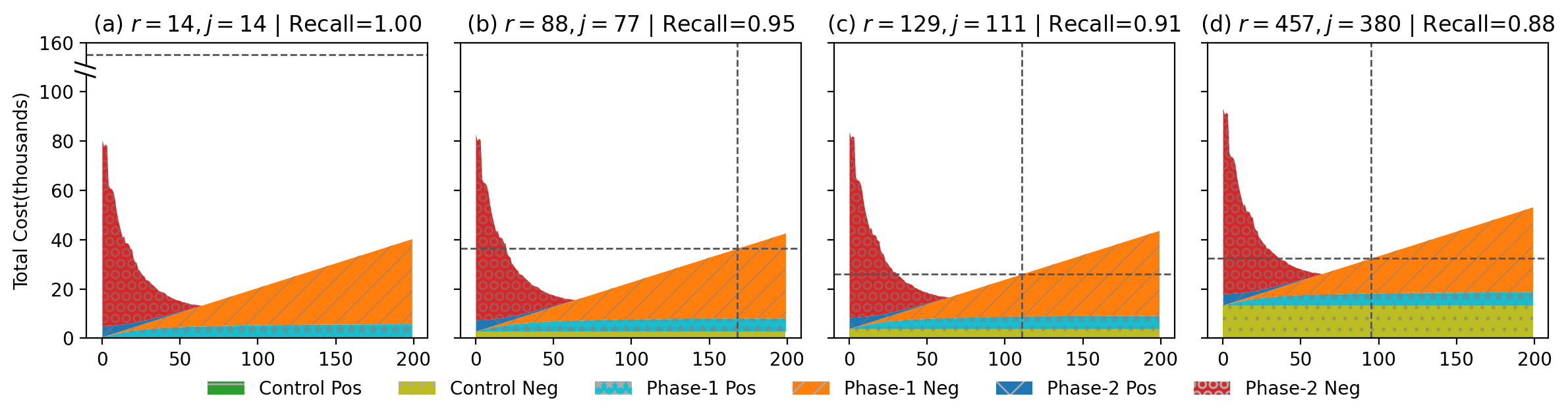}
    \caption{An example cost dynamics analysis.  For category  \texttt{E12} (common-hard bin) we show the components of total cost for each of the first 200 iterations of the TAR process. For iterations where recall 0.80 has not been reached, we add a penalty equal to the cost of an optimal Phase 2 continuation reaching 0.80 recall. For each of four sample sizes we choose the worst stopping point across 100 replications. Its cost is shown as a horizontal dashed line, and the stopping point as vertical dashed line. For sample size 14, worst case stopping is at iteration 759 with a cost of 152,324.}
    \label{fig:cost-dynamics}
\end{figure*}

Figure~\ref{fig:recall-boxplot} displays a boxplot of recall values at the stopping point for category E12 (from the common-hard bin) using 100 replications of each positive sample size and the QBCB rule. We see the usual decrease in variance with increased sample size. Only a few outliers are below a recall of 0.80 at any sample size.  Measures of central and high recall consistently decrease. 

We would expect that reducing the occurrence of very high recall values would also reduce the occurrence of very high costs.  Figure~\ref{fig:cost-boxplot} explores this in detail.  It is again a boxplot for 100 replications, but this time for all 9 of our exemplar TAR workflows and displaying total cost rather than recall. 

Category ALG (Algeria) is a category where Cormack and Grossman's approach of of using the minimum possible sample size leads to minimum cost.  The TAR workflow  reaches recall of 1.0 after only 32 batches, so overshooting recall can never cost the review of more than $32\times 200 = 6400$ documents. Conversely, the prevalence of positive examples is very low (0.002437)  so random positive examples are very expensive (each costing on average 410.34 negative examples).  

For most categories however, investing in random samples large enough to get more than the minimum number of positives brings down the maximum cost over 100 replications substantially. For I22100 (medium frequency and medium difficult) the maximum cost over 100 replications is a factor of 14 times greater for a sample of 14 positive than for an optimal sample of 30 positives. The graphs also emphasize the importance of a power analysis in choosing small sample sizes. For most categories and cost statistics, 21 positives is actually worse than 14, while 22 is better.

For categories E12 (Common-Hard) and I300003 (Common - Medium), using a larger than minimum sample size brings down not just the worst case cost, but even the median cost.  It is worth noting while these categories are in our "Common" bin, their prevalences are 3\% and 1\% respectively, which is typical or even low for e-discovery projects, depending on collection strategies. Sample-based stopping will be even more practical for a project in, say, the 10\% prevalence range.

\subsection{Cost Dynamics}

Our focus in this study has been on one-phase TAR reviews. Was anything lost by not considering two-phase review? Figure~\ref{fig:cost-dynamics} uses cost dynamics graphs~\cite{cost-structure-paper} to provide a perspective on this question. For a single TAR run (i.e., one seed) on category E12 we plot the total cost at stopping points from 0 to 200 iterations for four sample sizes. In addition to the four costs accounted for in Figure~\ref{fig:cost-boxplot}, for iterations where stopping would give recall less than 0.80 we add the cost of an optimal second phase review to reach 0.80 recall. That is, for each iteration we rank the unreviewed documents and assume that a top-down review through that ranking is carried out until 0.80 recall is reached. This is the minimum cost that a two-phase review reaching 0.80 recall would incur. 

The graphs immediately show that a one-phase review is optimal for this example: the minimum cost is at a point where no second-phase cost is incurred. This is typical for the setting of this paper, where the costs of all forms of review (sampling, phase 1, and phase 2 if present) are equal. One-phase review is typically not optimal when costs are unequal~\cite{cost-structure-paper}.  

The graphs also provide an interesting perspective on the role of sample size in minimizing cost. A horizontal dashed line shows the worst case total cost for QBCB over 100 replications for each sample size, while the vertical line shows the corresponding stopping point. As the sample size is increased, the stopping point comes closer to the minimum of the cost landscape, but the entire landscape is raised. The sample size that minimizes the worst case cost over our 100 replications, sample size 129 in this case, strikes a balance between the two effects.

\section{Future Work}

The QBCB rule makes use only of the positive documents in a random sample. Exploiting both positive and negative documents using a hypergeometric distribution should modestly reduce sample size, if the unknown number of 
relevant documents can be addressed. The bounding technique proposed in 
\citet{callaghan2020statistical} %
is one possible approach, as is treating the positive subpopulation size as a nuisance parameter \cite[Chapter~6]{lehmann2006theory}.  Other approaches to reducing sample size that could be applied are stratified sampling \cite[Chapter~11]{thompson1997theory} and multi-stage or sequential sampling \cite[Chapter~13]{thompson1997theory}. Dimm \cite{dimm2017confirming} has presented promising results on using multi-stage sampling to reduce costs in making a binary acceptance decision for a complete TAR production, and this approach likely can be adapted to stopping rules. 

Desirable extensions of QBCB would be to two-sided confidence intervals, to two-phase workflows \cite{cost-structure-paper}, to multiple assessors who may disagree, to effectiveness measures other than recall, and to rolling collections (where the TAR workflow must be started before all documents have arrived). Techniques from survey research for repeated sampling may be applicable to the last \cite{ohlsson1998sequential}.

Finally, the QPET and QBCB rules are based on viewing a one-phase TAR process as incrementally exposing a ranking of a collection. The rules may also be applied to actual rankings of collections produced by, for instance, search engines and text classifiers. In this scenario, QPET and QBCB become rules for avoiding sequential bias in choosing a sample-based cutoff that hits an estimated recall target.

\section{Summary}

The philosophy of statistical quality control is to accurately characterize and control a process \cite{grant1996statistical}. We have shown in this study that previously proposed certification rules for one-phase TAR reviews are statistically invalid, inflexible, expensive, or all three. 

Drawing on the statistical theory of quantile estimation, we derive a new rule, the QBCB rule, that avoids sequential bias and allows controlling the risk of excessive costs. The rule applies to any one-phase TAR workflow, and can immediately be put into practice in real-world TAR environments.  By using this rule, valid statistical guarantees of recall can be produced for the first time, while mitigating the risks of extreme cost. 

\begin{acks}
We thank Lilith Bat-Leah and William Webber for their thoughtful feedback on drafts of this paper, and Tony Dunnigan for the Figure 1 diagram. All errors are the responsibility of the authors.    
\end{acks}

\bibliographystyle{ACM-Reference-Format}
\bibliography{sample-base}

\end{document}